\documentclass[12pt,preprint]{aastex}
                                                                                
\shorttitle{Omgoing Formation of Bulges and Black Holes}
\shortauthors{Kauffmann et al.}
                                                                                
\begin{document}
                                                                                

\title{Ongoing Formation of Bulges and Black Holes in the Local Universe:
New Insights from GALEX}

\author{
Guinevere Kauffmann \altaffilmark{1},
Timothy M. Heckman\altaffilmark{2}, 
Tamas Budavari\altaffilmark{2}, 
Stephane Charlot \altaffilmark{3}, 
Charles G. Hoopes\altaffilmark{2}, 
D. Christopher Martin\altaffilmark{4}, 
Mark Seibert\altaffilmark{4}, 
Tom A. Barlow\altaffilmark{4},
Luciana Bianchi\altaffilmark{2}, 
Tim Conrow\altaffilmark{4},
Jose Donas\altaffilmark{5},
Karl Forster\altaffilmark{4}, 
Peter G. Friedman\altaffilmark{4},
Young-Wook Lee\altaffilmark{6},
Barry F. Madore\altaffilmark{7,8}, 
Bruno Milliard\altaffilmark{5},
Patrick F. Morrissey\altaffilmark{4}, 
Susan G. Neff\altaffilmark{9},
R. Michael Rich\altaffilmark{10}, 
David Schiminovich\altaffilmark{11}, 
Todd Small\altaffilmark{4},
Alex S. Szalay\altaffilmark{2},
Ted K. Wyder\altaffilmark{4},
S.K. Yi\altaffilmark{6}
}

\altaffiltext{1}{Max-Planck-Institut fur Astrophysik, Karl-Schwarzschild-Str. 1, D-85748 Garching bei Munchen, Germany}
\altaffiltext{2}{Department of Physics and Astronomy, Johns Hopkins University, 3400 N. Charles St., Baltimore, MD, 21218}
\altaffiltext{3}{Institut d'Astrophysique de Paris, UMR 7095, 98 bis Boulevard Arago, F-75014 Paris, France}
\altaffiltext{4}{Division of Physics, Mathematics, and Astronomy, California Institute of Technology, Pasadena, CA, 91125}
\altaffiltext{5}{Laboratoire d'Astrophysique de Marseille, BP 8, Traverse
du Siphon, 13376 Marseille Cedex 12, France}
\altaffiltext{6}{Center for Space Astrophysics, Yonsei University, Seoul
120-749, Korea}
\altaffiltext{7}{Observatories of the Carnegie Institution of Washington,
813 Santa Barbara St., Pasadena, CA 91101}
\altaffiltext{8}{NASA/IPAC Extragalactic Database, California Institute
of Technology, MC 100-22, 770 S. Wilson Ave., Pasadena, CA 91125}
\altaffiltext{9}{Laboratory for Astronomy and Solar Physics, NASA Goddard
Space Flight Center, Greenbelt, MD 20771}
\altaffiltext{10}{Department of Physics and Astronomy, University of California at Los Angeles, 8965 Mathematical Sciences Building, Los Angeles, CA 90095}
\altaffiltext{11}{Department of Astronomy, Columbia University, MC 2456, 550 West 120th Street, New York, NY 10027}

\email{gamk@mpa-garching.mpg.de}

 

\begin{abstract}
We analyze a volume-limited sample of massive bulge-dominated  galaxies with
data from both the Sloan Digital Sky Survey and the Galaxy Evolution
Explorer (GALEX) satellite.
The galaxies have central velocity dispersions greater than 100 km/s 
and  stellar surface  mass densities that lie above the value where
galaxies transition from actively star forming to passive systems.
The sample is limited to  redshifts $0.03 < z < 0.07$. At these
distances, the SDSS spectra sample the light from the bulge-dominated
central regions of the galaxies.
The GALEX  NUV data provide high sensitivity    
to low rates of {\em global} star formation in these systems. 
Our sample of bulge-dominated  galaxies exhibits a much larger
dispersion in NUV-$r$ colour than in optical $g-r$ colour. The  dispersion
increases for  galaxies with smaller central velocity dispersions and
nearly all of the galaxies with bluer NUV-$r$ colours are AGN.
Both  GALEX images  and  SDSS colour profiles demonstrate that
the excess UV light is nearly always associated with an extended disk. 
When comparing fibre-based estimates of stellar age
to global ones, we find that                
galaxies with red outer regions almost never have a young bulge
or a strong AGN. Galaxies with blue outer regions have bulges and black holes
that span a wide range in age and accretion rate. Galaxies with
young bulges and strongly accreting black holes almost {\em always} have blue
outer disks.  The black hole growth rate correlates much more strongly
with the age of the stars in the bulge than in the disk. 
Our suggested scenario is one in which the source of gas that builds the
bulge and black hole is a low mass reservoir of cold gas in the disk. The
presence of this gas is a necessary, but not sufficient
condition for bulge and black hole growth.
Some mechanism must transport this gas inwards in a time variable way. 
The disk gas itself is likely to be the result of the accretion of gas
from an external source. 
As the gas in the disk is converted into stars, galaxies will turn
red, but  further inflow can  bring them back into the blue NUV-$r$
sequence. 
\end{abstract}
 
 

 

\section {Introduction}

The formation and evolutionary history of massive  bulge-dominated
galaxies has been a subject of considerable controversy over the
past decade. On the one hand, the colours and detailed 
spectral energy distributions of these galaxies indicate that their
stellar populations are predominantly old and metal-rich (see Renzini (2006)
for a recent review).
The dispersion in spectral properties between different systems is
small, indicating that the stars in different galaxies were  
formed at  roughly similar epochs and less than a few gigayears
after the Big Bang (e.g. Bower, Lucey \& Ellis 1992)
On the other hand, a theoretical paradigm
for the assembly of structure in the Universe has emerged in
recent years, which at first sight has appeared to contradict
this observational data. According to this paradigm, the largest
structures in the Universe form {\em hierarchically} through
the merging of small virialized condensations (halos) to form larger
and larger systems. This means that the dark matter halos
that host massive bulge-dominated galaxies are predicted to have  
assembled relatively recently. In addition, a significant fraction of 
these galaxies  are expected to reside at the centers of 
massive halos, where gas is expected to reach high enough
densities to be able to cool through radiative processes. 
The predicted cooling rates in these
halos are high 
enough to produce a population of young stars that should  be  
easily detectable in the spectra of their central galaxies
(see for example Kauffmann, White \& Guiderdoni 1993).

In order to resolve this so-called "cooling crisis", a number
of recent models invoke feedback from the active galactic nuclei
in massive galaxies. Energy from the material accreting onto
the central supermassive black hole is transported to the gas
surrounding the galaxy either by jets that are known to exist in radio
galaxies (e.g. Churazov et al 2001; Ruszkowski et al 2004; Best et al 2005; 
Croton et al 2006) or by powerful outflows that are hypothesized to be
triggered during galaxy-galaxy mergers 
(Granato et al 2004; Di Matteo, Springel, Hernquist 2005). 

All these models are highly schematic, so it is
valuable to  use observational data to provide a 
more detailed understanding of the
cooling, star formation and  heating processes that regulate
the evolution of massive galaxies. 
Although the present-day  star formation rates in massive galaxies
galaxies are constrained to be low on average, 
there is observational evidence that stars
do continue to form in a subset of these systems. The optical
spectra of central cluster galaxies frequently exhibit line                
emission and blue continua (McNamara \& O'Connell 1989; 
Cardiel, Gorgas \& Aragon-Salamanca 1998; Crawford et al 1999). 
Excess ultraviolet light has also been detected in number of central
cluster galaxies (Hicks \& Mushotzky 2005).
The inferred star formation rates in these objects are similar  to
rate at which gas is measured  to be cooling
at the very centers of the clusters using X-ray data from
the Chandra and XMM satellites (McNamara 2004).

Recently, near and far-ultraviolet photometry from the
Galaxy Evolution Explorer (GALEX) has been assembled for a large sample of
galaxies from the Sloan Digital Sky Survey.    
The advantage of studying galaxies at ultraviolet wavelengths is that their    
UV luminosities are very sensitive to low levels of residual star formation.
The disadvantage is that rest-frame ultraviolet data can
only be obtained from space for galaxies at low redshifts, so until recently
the available galaxy samples have been small.  
Yi et al. (2005) and Rich et al. (2005) presented analyses of the
UV/optical color magnitude relation of two samples of early-type galaxies
selected from the first matched GALEX/SDSS sample. These studies found that
the UV/optical colors of these galaxies exhibit significantly larger scatter
than their optical colors. The main conclusion was that low-level residual
star formation was common during the last billion years, even in bright
early-type galaxies. Rich et al. (2005) also noted that early-type galaxies
with AGN were much more likely than quiescent galaxies to show blue
UV/optical colors. 
Possible implications of the above results for models of elliptical
galaxy formation have been
considered in recent papers by  Kaviraj et al. (2005;2006), 
Martin et al (2006)  and Schawinski et
al. (2006a,b). 

We note, however, that Boselli et al (2005) carried out an 
analysis of UV properties of a sample of  nearby elliptical galaxies  in
the  Virgo cluster, and found rather tight colour-magnitude relations 
for the objects in their sample. The ellipticals in this sample were selected
using high quality photographic plates  taken with the DuPont telescope
at Las Campanas Observatory. The differences between the Boselli et al analysis
and the studies that use the SDSS imaging data to select early-type galaxies, suggest
that UV properties of early-type galaxies may be quite  sensitive to
the exact selection criteria used to define the different samples.       

In this paper, we adopt a different approach.                   
Using a  parent sample of galaxies with GALEX detections that
are matched to the SDSS Data Release 4 (DR4)  sample, 
we have constructed a volume-limited sample of bulge-dominated
galaxies in the local Universe with near-UV and optical photometry,
as well as optical spectroscopy from the SDSS. Unlike the previous
studies, our sample is selected on the basis of  bulge velocity dispersion
and {\em not} on the basis of conventional Hubble type. Our sample is also limited to 
lie at redshifts less than 0.07. At these distances the galaxies
subtend a large enough solid angle on the sky to enable us to
study the {\em radial distributions} of the young stars in the galaxies using
both the GALEX and the SDSS images. In addition, the spectra of these
galaxies have very high signal-to-noise, enabling us to accurately
deconvolve even intrinsically weak emission lines from the underlying
stellar absorption continuum and to study the properties of AGN in
these objects.  In section 2, we describe the details of our sample
selection. In section 3, we compare the UV/optical colour relations         
of bulge-dominated galaxies with the ``traditional'' relations defined
only in the optical. In section 4, we analyze the spatial distribution
of the young stars in galaxies with blue NUV-$r$ colours. In section 5,
we examine the link between low levels of
ongoing star formation and AGN activity in our sample. 
In section 6, we study the stellar mass profiles of the galaxies 
in our sample. Finally in section 7,
we summarize and interpret our results.

\section{The Sample}

\subsection{Ultraviolet Data}

Since its launch in April, 2003, GALEX has been conducting several
surveys of the UV sky. 
 Details on the GALEX mission and surveys
are given in Martin et al (2005) .
In this paper we make use of the GALEX                  
Medium Deep Survey (MIS). MIS exposures are typically 1500 seconds long 
and reach a limiting magnitude $m_{AB} \sim 23$. 
Our data are taken from the 
GALEX MIS IR1.1(+GR1) data release.
The GALEX data include far-ultraviolet (FUV; $\lambda_{eff}=1528$~\AA,
$\Delta\lambda=268$~\AA) and near-ultraviolet (NUV;
$\lambda_{eff}=2271$~\AA, $\Delta\lambda=732$~\AA) images with a
circular field of view with radius $\sim 38$ arcminutes. The spatial
resolution is $\sim 5$ arcseconds. Details of the GALEX satellite and data
characteristics can be found in Morrisey et al (2005). 
The data were processed through the GALEX reduction pipeline at the
California Institute of Technology. The pipeline reduces the data and
automatically detects, measures, and produces catalogs of FUV and NUV
fluxes for sources in the GALEX images.

\subsection{Optical data}

The GALEX catalogs were then matched to the SDSS Fourth Data Release
(DR4; Adelman-McCarthy et al. 2006) spectroscopic sample. 
The matching procedure is described in Seibert et al (2005). 
Matching to SDSS provides
a variety of photometric parameters in the $u,b,v,r,i,z$ photometric passbands as well
as spectroscopic redshifts for a sample of 51,246 galaxies with UV detections. 
In this analysis, we  exclude galaxies located more than 30 arcminutes 
from the MIS field centers , because the photometry is not as reliable
near the boundaries.

In addition to the SDSS photometric parameters, 
a large number of galaxy parameters derived from the SDSS spectra are
available in the value-added catalogs at the SDSS website at the 
Max Planck Institute (http://www.mpa-garching.mpg.de/SDSS). From these
catalogs we use the stellar masses, stellar velocity dispersions,
the spectral indices D$_n$(4000) and H$\delta_A$, the equivalent
width of the H$\alpha$ emission line (EQW(H$\alpha$)), and the AGN
classifiers. For detailed information about the derivation of these
quantities , see Kauffmann et al (2003a,c); Brinchmann et al (2004)
and Tremonti et al (2004). 
Throughout this paper we have assumed   $H_0=70$~km~s$^{-1}$ Mpc$^{-1}$,
$\Omega_m=0.3$, and $\Omega_{\Lambda}=0.7$.

The aim of the present analysis  is to study a volume-limited sample of
bulge-dominated galaxies. It is thus important to make sure that our sample,
which is initially selected in the UV, is not biased {\em against}  galaxies
with no ongoing star formation. In Figure 1, we show how the NUV-$r$ and
$g-r$ colours of galaxies depend on the fraction of the stellar mass
that was formed in the past Gyr. These curves were generated using
the Bruzual \& Charlot (2003) GALAXEV code assuming a solar metallicity  
stellar population and smooth, exponentially declining star formation
histories with varying e-folding times. As can be seen from the
plot, as the fraction of young stars in the galaxy goes to zero,
the NUV-$r$ colours are predicted to saturate at a value of around 6.5.    
The figure also shows that the NUV-$r$ colour remains sensitive to 
a much lower level of recent star formation than the optical $g-r$ colour.
Since our GALEX sample is limited at $m_{AB} =23$, Figure 1 suggests that 
an $r$-band magnitude limit of 16.5 (i.e. 1.2 magnitudes brighter than
the limit of the main spectroscopic sample) is required in order to
ensure that our sample is not biased against galaxies with
the very oldest stellar populations.

The top left panel in Figure 2 shows the NUV-$r$ colour distribution of
galaxies with $r< 16.5$ and NUV$< 23$ in our sample. We use SDSS
model magnitudes, which are well matched to the AUTO/Kron  magnitudes
output by the GALEX pipeline. The magnitudes have also been
corrected for foreground extinction using the values of E(B-V)
given by Schlegel et al (1998) and the Cardelli et al (1998) reddening curve.  
As can be seen, the colour
distribution is clearly bimodal. The population of red galaxies
is peaked at NUV-$r$ $\sim 6$ and there are very few  galaxies   
with NUV-$r$$> 6.5$. This gives us confidence that our sample is indeed
complete in NUV-$r$ colour space. In the next two panels, we plot the
stellar masses and the stellar velocity dispersions (measured within
the 3 arcsec fiber aperture) of the red galaxies with
NUV-$r$ $>4.5$  as a function of their redshifts. These red objects
are the most difficult to detect in the UV and they also have the highest
$r$-band stellar mass-to-light ratios, so they determine the
effective completeness limit of our sample.
Figure 2 shows that if  we limit the sample to lie in the 
redshift range $0.03 < z < 0.07$, we are complete
down to $\log M_* \sim 10.4$ and $\log \sigma \sim 2.05$.

We note that at $z \sim 0.05$, the SDSS fibre aperture subtends a
physical scale of only 3.2 kpc. The physical quantities that are
measured from the spectra are thus relevant to the {\em central bulge-dominated
regions of the galaxy}. Throughout this paper, we will interpret spectral
quantities such as D$_n$(4000), H$\delta_A$ and EQW(H$\alpha$) as measures of the
age of the stellar population {\em in the bulge}. The NUV-$r$ and
$g-r$ colours , on the other hand, are sensitive to the age of the stellar population
of the galaxy as a whole.

In the bottom right panel of Figure 2 we plot the stellar surface
mass density $\mu_*$  as a function of stellar velocity dispersion 
for galaxies with $\log M_* > 10.4$, $\log \sigma > 2.05$,  $r< 16.5$,
NUV$<23$ and $0.03 < z < 0.07$ . This sample includes 1375 galaxies. 
The surface mass density $\mu_*$ is defined as $\mu_* = (0.5 M_*)/(\pi R50z)^2$,
where $R50z$ is the half-light radius of the galaxy in $z$-band.
As discussed in Kauffmann et al (2003b; 2006), there is a sharp
transition in the mean stellar age of galaxies at  a stellar
surface mass density $\mu_*$ of around $10^{8.5} M_{\odot}$ kpc$^{-2}$.
Galaxies with surface densities lower than this value have 
on average formed their stars at a constant rate for a Hubble time;
galaxies with surface densities higher than this value have little or
no ongoing star formation. Figure 2 shows that essentially all
galaxies in our sample defined at $\log \sigma > 2.05$ have surface
densities greater than the transition value. The range of densities spanned by
the objects in our sample is also very similar to that of elliptical galaxies
(Bernardi et al 2003)  and  it is hence meaningful to
refer to these objects as ``bulge-dominated'' systems. 

We also find that 85\% of the galaxies in our sample 
have $r$-band  concentration indices
(C=R90/R50) greater than 2.6. Strateva et al (2001) and Shimasaku
et al (2001) have studied how this index relates to "by eye" classification
into Hubble type. These authors find that C=2.6 marks a reasonably robust
division between early-type galaxies (E/S0/Sa) and the late-type
spiral population. Note that we choose not to apply further 
cuts to exclude the minority of  later-type galaxies in our sample.
The aim of the present paper is to examine possible evolutionary links between
different galaxy populations  at fixed bulge velocity dispersion
or mass. It is thus critical to ensure that  our sample is  {\em complete} at
a given value of $\sigma$.

\section{Colour Relations}

In Figure 3, we plot the NUV-$r$ and $g-r$ colours
of the galaxies in our sample as a function of the
stellar velocity dispersion measured within the SDSS fiber. 
The points have been colour-coded according
the emission line properties of the galaxies. Galaxies with emission
lines that are too weak to permit a secure classification are coloured
in black, AGN are coloured in red and star-forming galaxies are coloured
in blue. As can be seen, the scatter in the NUV-$r$/$\sigma$ relation
is substantially larger than the scatter in the $g-r$/$\sigma$
relation. The scatter increases at lower values of $\sigma$ in both
diagrams. It is striking that at velocity dispersions greater than
$\sim$ 125 km s$^{-1}$, almost all the galaxies with blue NUV-$r$ colours  
are classified as AGN. There are only a small number of galaxies
with emission line ratios typical of star-forming galaxies and 
these all have small central velocity dispersions. These results are
quantified in more detail in Figure 4, where we plot colour distributions
in different ranges of stellar velocity dispersion. The black histograms
represent the results for the full sample, while the red and blue 
histograms show the contributions from AGN and star-forming galaxies
respectively.

Figure 5 examines the relations between four different stellar population
indicators: the NUV-$r$ colour, the $g-r$ color,
the 4000 \AA\ break index D$_n$(4000),  and the equivalent
width of the H$\alpha$ emission line. 
As discussed in the previous section, the NUV-$r$ and
$g-r$ colors are measures of the stellar populations of
the galaxy as a whole, whereas D$_n$(4000) and EQW(H$\alpha$)
probe the stellar populations within the central bulge-dominated
region (Note that for galaxies classified as AGN, the H$\alpha$
will be sensitive to the ionizing flux from
the central source as well as the HII regions in the galaxy.) 
Figure 5 shows that there is a tight, but strongly  
non-linear correlation between the $g-r$ and NUV-$r$ colours.
As shown in Figure 1, galaxies with NUV-$r$ colours in the 
range 4 to 6 have formed less than 0.3 percent of their stars over
the last Gyr. The $g-r$ optical colours are completely insensitive
to such low levels of star formation; this is why $g-r$
is flat over this range in NUV-$r$ colour. The two spectral 
indicators measured within the fibre (D$_n$(4000) and
EQW(H$\alpha$))  also exhibit a tight correlation. At fixed D$_n$(4000),
AGN (red) scatter to slightly higher H$\alpha$ equivalent widths than
star-forming galaxies (blue), consistent with the notion
that radiation from the active nucleus contributes to
the ionization of the H$\alpha$ line. Nevertheless the relation
between the two indicators is still quite tight for both classes of object. 
However, when 
we plot the fibre indicators as a function of the global
NUV-$r$ colours, we see a huge scatter. Galaxies with NUV-$r$
colours in the range $2-4$  have H$\alpha$ equivalent widths that differ 
by more than an order of magnitude, and D$_n$(4000) values that span  
the entire dynamic range of the index.

The most obvious explanation for this enormous scatter 
is that the {\em UV light does not trace
the stellar population in the bulge.} As we have noted previously,
our sample is chosen to lie at redshifts below 0.07 and
the physical aperture subtended by the SDSS fibre is only a few 
kiloparsecs.  The hypothesis that the UV light does
not originate from the bulge is consistent with
recent GALEX studies of very nearby galaxies, which indicate  that a 
significant fraction of the UV light originates in the outer disks
of galaxies (Thilker et al 2005). Popescu et al (2005) show that
the UV-to-far infrared flux ratio in the galaxy M101 is a strong function
of radius, with values monotonically decreasing from $\sim 4$
in the nuclear region to nearly zero at the edge of the optical disk. 
In the next section, we will prove that the UV light is biased
towards the outer regions of the galaxies in our sample
by analyzing their mean optical radial color profiles
as a function of their global  NUV-$r$ colours.

The SDSS photometric pipeline extracts azimuthally-averaged
radial surface brightness profiles for all the objects
in the survey. In the catalogues, this  
is given as the average surface brightness 
in a series of annuli with fixed angular dimension.
Results are stored in the $u,g,r,i,$ and $z$-bands, so it
is simple to compute radial colour gradients for the
galaxies in our sample. In figure 9, we plot $g-i$ as a function
of physical radius for UV-bright (upper panels) and UV-faint
(lower panels) galaxies with stellar velocity dispersions in
the range $2.05 < \log \sigma < 2.25$. The average profile
is shown as a solid line; the dashed lines indicate the 10-90
percentiles in the range of colour spanned by
the galaxies at a given radius.
The  colour profiles are  smooth because we  average
over samples of between 100 and 300 objects. We only plot
results out to radii where the photometric errors in the
flux contained within the radial bin are less than $\sim 10$\%.

It is clear from Figure 9 that UV-bright galaxies have very similar
central colours to UV-faint galaxies, but are significantly bluer
in their outer regions. UV-bright galaxies with strong
emission lines in their central regions exhibit a larger
scatter in their central $g-i$ colours, but  their
{\em average} nuclear colours are  very similar to their UV-bright counterparts
with weak emission lines. As was noted in the previous section,  
essentially all UV-bright galaxies with central stellar velocity
dispersions in this range are classified as AGN. In the bottom
two panels, we compare the colour profiles of UV-faint galaxies
that are classified as AGN with those with emission lines
that are too weak to classify. The  colour profiles
of the two kinds of galaxy are very similar.  UV-faint AGN are 
slightly bluer in their outer regions on average.

\section{Colour Profiles}

In Figures 6 and 7, we show a montage of SDSS $g,r,i$
and GALEX FUV, NUV colour images of bulge-dominated galaxies 
with blue NUV-$r$ colours (NUV-$r< 4.5$). In Figure 8 we show
a sample of bulge-dominated galaxies with red
NUV-$r$ colours (NUV-$r > 5$) for comparison.   All the galaxies
are selected to lie  between redshifts 0.04 and 0.06,                    
and they all have central stellar velocity dispersions in the
range $2.1<\log \sigma<2.4$. Each postage stamp image
is 100 arcsec on a side, which corresponds to a physical
scale of $\sim$100 kpc. In Figure 6, we show galaxies
with NUV-$r<4.5$ and low H$\alpha$ equivalent widths measured within the fibre 
( EQW H$\alpha$ $> -2$ \AA\ ). Figure 7 shows galaxies with 
NUV-$r < 4.5$ and EQW H$\alpha$ $< -7$ \AA (note that essentially all of
these are AGN). There is no explicit
emission line cut in Figure 8, but as shown in Figure 5, almost all 
galaxies with red NUV-$r$ colours have low emission line equivalent widths. 

The SDSS colour images have been produced using an algorithm described in
Lupton et al (2004).  The algorithm uses an 
``arcsinh stretch'' that allows faint,
low-surface brightness features in a galaxy  to be displayed, while
simultaneously preserving the structure of the 
bright, high-surface brightness components.  Even at recessional
velocity of 15,000 km/s, these galaxies display a rich
complexity of structure in the SDSS images, with outer rings, bars and spiral     
structure clearly visible in many objects. \footnote {Note  that
if the same galaxies are viewed at larger distances, this structure
becomes increasingly difficult to see in the SDSS postage
stamp images. We inspected  SDSS images of galaxies with the same range
of central velocity dispersions at $z=0.1$
and we find that although outer disks are sometimes visible, 
it is no longer possible to discern star-forming regions in the images.}
The GALEX UV images
have considerably lower spatial resolution than the SDSS
images, but they  make it clear that the UV light is
extended over large spatial scales. In Figure 6,
there are galaxies where the UV appears to extend  beyond 
the optical radius of the galaxy, out to galactic radii as large
as 40-50 kpc.  The UV-bright  galaxies displayed  in Figure 7 have higher UV
surface brightnesses than those in Figure 6, but the UV 
emission still clearly extends over the entire  region
traced by the optical light, and sometimes beyond. On the other hand the  UV-faint
galaxies displayed in Figure 8 occasionally display some extended emission, 
but in  most cases the UV light appears to be concentrated in the same  
bulge-dominated regions of the galaxies as the optical light.

\section{What is the link between the UV emission and the AGN activity?}

In section 3, we showed that bulge-dominated galaxies with blue NUV-$r$
colours almost always have central emission line spectra characteristic
of ionization by an active nucleus. In section 4, 
we showed that the UV emission in bulge-dominated galaxies 
is a tracer of young stars in the {\em outer} region of the galaxy.
This leads us to the following question: what, if any, is the
physical connection between this extended UV emission and the 
activity in the nucleus of the galaxy?

In order to address this issue, it is necessary to move beyond
simple classification of the galaxies in our sample into AGN and   
and non-AGN. Heckman et al (2004) introduced the [OIII] emission line 
luminosity as an indicator of the rate at which matter is accreting onto
the central supermassive black hole. There are two 
reasons why the [OIII] emission line luminosity
is believed to be a good indicator of accretion rate: 
1) [OIII] emission  is relatively weak
in metal-rich HII regions; 2) In type I Seyfert galaxies and quasars, 
the [OIII] line luminosity is well correlated with the continuum luminosity and 
by extension, the black hole accretion rate. Assuming that
the Unified Model is valid, the same should hold true in galaxies
where the central engine is obscured. In follow-up work, Heckman et al (2005)
studied a sample of hard X-ray selected AGN, and showed that the
hard X-ray emission, which is believed to be an unbiased
tracer of accretion on to the black hole, and the [OIII] emission-line 
luminosities were well-correlated over a range of 4 orders of magnitudes in 
luminosity in both Type I and Type II AGN. 
Since we have measurements of the  central stellar velocity dispersions
for all the galaxies in our sample, we are able to estimate black hole masses
using the relations given in Tremaine et al (2002). The ratio 
L[OIII]/M$_{BH}$  is then a measure of the accretion rate relative
to the Eddington rate. 

We note that the black hole masses and accretion rates estimated from the
SDSS spectra should be regarded as indicative rather than precise measures 
of these quantities.    
The stellar velocity dispersions in Tremaine et al. (2002) were measured within
phyical apertures that were on average a factor of 4 smaller than those 
corresponding to the 1.5 arsec fiber diameter for the galaxies in our sample.
As discussed in Heckman et al (2005), there is at least a factor of
a few uncertainty in the conversion from an [OIII] emission line luminosity
to an accretion rate. Nevertheless, as we will demonstrate, the
stellar populations of the galaxies in our sample correlate remarkably  tightly
with L[OIII]/M$_{BH}$.

Less than 15 \% of the SDSS
sample is covered by GALEX observations. However, we have seen that even if we
lack UV data, we can still use the SDSS $g-i$ colour profiles                    
to see whether there are young stars in the outer regions of our galaxies.
We have thus chosen to stack galaxies with similar central velocity
dispersions in bins of L[OIII]/M$_{BH}$ and to study how their optical
colour profiles change as a function of black hole accretion rate.
This is shown in Figure 10 for galaxies with
$2.05 < \log \sigma < 2.25$. Solid cyan, blue, green, red, magenta and black 
curves show the average $g-i$ colour profiles of galaxies hosting AGN
with decreasing values of L[OIII]/M$_{BH}$. \footnote {Note that
we use extinction corrected values of L[OIII] throughout this paper.} 
The dashed and dotted curves  
show the upper 90th and lower 10th percentiles of the distribution
of $g-i$ colour at a given radius. The striking result indicated in this
figure is that AGN with higher accretion rates have bluer colours
in their outer regions. In contrast, the average colour
in the central regions of the galaxy appears to exhibit  
no dependence on the accretion rate onto the central black hole.

At first sight, this might appear to be a paradoxical result.
It would be difficult to come up with any physically motivated scenario 
that would predict the accretion onto the central black hole to be  
more strongly modulated by conditions in the outer
rather than the inner regions of the galaxy!
It is important to remember, however,  that galaxy colours are
sensitive not only to the ages and metallicities of their stellar
populations, but also to the amount of dust present
in the interstellar medium. Within the aperture covered 
by the SDSS fibre aperture, information available from the SDSS
spectra allow us to probe mean stellar age and the amount of dust  
in an independent way. This is illustrated in Figure 11.
In the upper two panels we plot the distribution of the two
age-sensitive indices D$_n$(4000) and H$\delta_A$ for galaxies
in the same ranges in L[OIII]/M$_{BH}$ as shown in Figure 10.
Here we see a very clear progression in the mean value of these
indices as a function of black hole accretion rate. The stellar
populations in the central regions of galaxies with strongly
accreting black holes are clearly younger. In the bottom panels
we plot the distribution of $A_z$, the attenuation of starlight
due to dust in the $z$-band (see Kauffmann et al (2003a)
for more details),  and the  Balmer decrement 
H$\alpha$/H$\beta$, which measures the extinction in regions
of the galaxy with young, massive stars. Both indicators 
show that there is more dust in the central regions of galaxies
with strongly accreting black holes. We conclude
that the reason why the average colours in the central regions of AGN
are so weakly correlated with accretion rate, is because
the bluing that would be expected as a consequence of the decrease in
the mean stellar age of the stars in the bulge is cancelled by the reddening 
due to the increase in the amount of dust. 
The {\em exactness} of this cancellation in the central bulge-dominated
region of the galaxy is remarkable and is worthy of further consideration.
In the outer (presumably disk-dominated) regions of the galaxy, age and dust  
effects no longer compensate each other exactly and as a result, these 
regions of the galaxy do become  bluer as the age of the stellar
population decreases. 

The results presented in Figures 10 and  11  indicate
that accretion rates onto central supermassive black holes
respond to conditions in {\em both} the outer and the inner regions
of the host galaxies. We now ask whether we can
ascertain  which response is the stronger of the two.
To answer this question, we divide our sample into bins in 
outer $g-i$ colour (measured at radii between 7 and 10 kpc)
and in inner stellar population age as measured by the H$\delta_A$
index. We study how the distribution in L[OIII]/M$_{BH}$ changes
as a function of these measures of the age of the  outer and the inner
stellar populations. The results are shown in Figures 12 and 13.
It is clear from these plots that the accretion rates respond
much more strongly to the age of the inner stellar population.

This conclusion is re-iterated in Figure 14, where we present 
the correlations between NUV-$r$ colour, D$_n$(4000) and   L[OIII]/M$_{BH}$
for galaxies with $2.05 < \log \sigma < 2.25$.
As we have discussed , the NUV-$r$ colour is sensitive to the 
age of the stellar population in the outer galaxy (effects due to
dust are expected to be weak), 
D$_n$(4000)  measures  the age of the stars in the inner bulge-dominated
region , and
L[OIII]/$M_{BH}$ measures the accretion rate onto the black 
hole. As can be seen, the black hole accretion
rate is much more tightly correlated with the mean stellar age in the bulge
than in the outer galaxy.
The black hole/outer galaxy and bulge/outer galaxy
correlations both have a  triangular shape.
This triangular correlation
tells us that galaxies with  high black hole accretion rates and young bulges                  
almost always have young stars in their outer regions.  However, a young outer region
{\em does not guarantee} that the bulge is young or the black
hole accretion rate is  high.

\section{Stellar Mass Profiles}

The most straightforward interpretation of the results presented in the
previous sections is that the UV light traces an extended reservoir of
HI gas that surrounds a subset of the bulge-dominated
galaxies in our sample. This outer gas is presumably in the form
of a rotationally supported disk and it  
contains the fuel that is required for further growth
of the the central supermassive black hole and the surrounding bulge.
The triangular shape of the correlations between
D$_n$(4000),  L[OIII]/$M_{BH}$ and NUV-$r$ colour  presented in
the first two panels of  Figure 11 implies  
that the  presence of gas in this disk is a 
{\em necessary, but not sufficient condition
for further bulge and  black hole growth.}  Black holes with low
accretion rates are found both in galaxies with old bulges and red      
NUV-$r$ colours (these are galaxies that are nearly devoid of gas)  and
in galaxies with old bulges and  blue NUV-$r$ colours 
(these galaxies do  contain gas, but in the form of an extended disk).
On the other hand, the tight correlation between 
L[OIII]/M$_{BH}$ and D$_n$(4000) presented in the third  panel of Figure 14
implies that the presence of  young stars (and presumably gas) in the bulge
is a {\em necessary and sufficient condition} for further black hole
growth. 
The strong correlation between black hole accretion rate
and the amount of extinction measured within
the SDSS fiber aperture (bottom panels of
Figure 11) is also clear evidence that black hole fuelling 
is strongly  linked to the amount of cold gas in the
inner region of the galaxy.
We propose that strongly accreting black holes are only found
in galaxies where the gas in the disk
has become more centrally concentrated, perhaps as a result of
angular momentum loss during an interaction. 

Unfortunately, HI data is not available for the galaxies
in our sample. However, we can test the hypothesis that the
extended UV-bright components shown in Figures 6 and 7
correspond to an outer disk by analyzing the
radial distribution of the {\em stars} in these galaxies.
To do this we need to transform the radial surface brightness profile
of the galaxy into a stellar mass profile. Bell \& de Jong (2001)
used spiral galaxy evolution models
to predict that the optical colours of galaxies are expected to be strongly
correlated with their stellar mass-to-light ratios. Both                            
stellar age and dust attenuation  affect 
the optical colours, but in such a way that 
$M/L$ can still be accurately predicted, provided that
the recent star formation history of the galaxy  has been reasonably smooth.
We have estimated stellar
mass-to-light ratios  within the SDSS fibre aperture
using a method  based on stellar absorption-line indices (Kauffmann et al 2003a).
We have correlated our estimates of $M/L$ with a variety of optical colours
and we find the tightest relation with the $g-i$ optical colour.
This is shown in  Figure 15. The black points
show our  SDSS/GALEX sample of bulge-dominated galaxies, while 
the blue points are for galaxies with smaller stellar masses and velocity
dispersions , but in the  same redshift range as the
galaxies in our bulge-dominated sample  ($0.03<z<0.07$). 
Note that we have taken care to plot the mass-to-light ratio in the $i$-band
as a function of the $g-i$ fiber colour, so that all quantities
are measured and evaluated  within the same physical aperture. 
As can be seen $M/L$ is well-correlated with $g-i$ fiber colour over a range
of more than a factor 10 in mass-to-light ratio 
and with an r.m.s. scatter of less than 0.1 dex.

This gives us confidence that we can use the $i$-band surface brightness
profile of a galaxy in conjunction with its  $g-i$ radial colour profile
to calculate the fraction of its stellar mass exterior to
a given radius in a galaxy. The results of this
calculation  are shown in Figure 16. 
We stack  galaxies within a narrow range in central stellar velocity disperion 
($2.05 < \log \sigma < 2.25$) and study whether  their stellar mass
profiles differ according to whether they are UV-bright or UV-faint.
In the left panel, solid black and red
curves show average mass profiles for UV-faint galaxies, 
while solid blue and green curves are for
UV-bright galaxies. The green curve shows the mass profile for UV-bright
galaxies with weak central emission line strengths, while the
blue curve is for UV-bright galaxies with strong central
emission lines. (Note that the colour-coding is the same
as in Figure 9).  As can be seen,
the stellar mass profiles of the UV-faint galaxies and the UV-bright
galaxies with strong central emission line strengths are very similar.
Only  UV-bright galaxies with weak central emission lines have  significantly               
different mass profiles; these galaxies have on average  10\% more mass in their
outer regions.

In the right panel of Figure 16, we plot the stellar mass profiles of
AGN ordered according black hole accretion rate. The colour-coding is
the same as in Figure 10. There is a  trend for galaxies with
higher black hole accretion  rates to have more concentrated stellar mass
profiles, but it is quite weak. Only the galaxies with black holes
that are accreting near the Eddington rate appear significantly more concentrated.
Note that these strongly accreting systems also have total stellar masses
that are about 10-20\% smaller than the galaxies with weakly accreting black
holes. The small difference in total
stellar mass at fixed value of the central velocity dispersion $\sigma$ is 
consistent with the notion that these galaxies all have very similar
total masses, but that the more strongly accreting objects have
slightly  higher gas fractions.

The evolutionary scenario suggested by the results presented in this section
is one in which bulges and their central supermassive black holes
form from gas located in an outer disk. 
These disks have presumably accreted 
from gas in the surrounding dark matter halo. So long as the galaxy
remains undisturbed, very little of the gas from this disk
will manage to reach the bulge and fuel an AGN. These quiescent
bulges/black holes with blue extended disks make up the
left bottom corner of the triangular cloud of points shown
in the first two panels of Figure 14. Eventually, some process
will lead to a flow of gas from the disk to the bulge.
This triggers star formation in the bulge and significant growth
of the central black hole. During this process, the increase of stellar mass in the
bulge is larger than the increase in the disk, so the galaxy develops
a more centrally concentrated stellar mass profile. 
Once the gas in the disk is exhausted, the mass profile of the
galaxy is indistinguishable from that of a classical elliptical galaxy.

\section {Summary and Conclusions}

We have defined a sample of massive bulge-dominated galaxies 
with both GALEX photometry and spectroscopic 
data from the SDSS. Our choice of limiting $r$-band 
magnitude ensures  that that the sample
is complete in the near ultra-violet pass-band. We have also imposed stellar
mass and redshift cuts to ensure that the sample is volume-limited for all
galaxies above $10^{10.4} M_{\odot}$ irrespective of their 
past star formation histories and present-day stellar mass-to-light
ratios. We study galaxies with central velocity dispersions greater
than 100 km/s. We have shown that these are bulge-dominated systems 
with  stellar surface  mass densities well above the value where
galaxies transition from actively star forming to ``passive'' systems.
The NUV data provide significantly higher sensitivity 
to low rates of star formation  in bulge-dominated galaxies. Our primary
observational results can be summarized as follows:

\begin {itemize}

\item  Bulge dominated galaxies exhibit a much larger
spread in NUV-$r$ colour than in optical $g-r$ colour. The  dispersion in
colour decreases for galaxies with larger central velocity dispersions.

\item Nearly all of the galaxies with blue NUV-$r$ colours are AGN.

\item  GALEX images  and  SDSS color profiles demonstrate that
the UV excess is associated primarily with an extended outer 
component of the galaxy.

\item  When comparing fiber-based properties to global ones, we find  ``triangular''
correlations. Galaxies with red outer regions almost never have a strong
AGN or a young bulge. Galaxies with blue outer regions have a wide
range in bulge/black hole properties. Galaxies with strongly accreting black
holes and  young bulges almost always have blue
outer regions. 

\item  The black hole growth rate correlates much more strongly with the age 
of the stellar population in the bulge than in the outer region of the galaxy.

\item 
The amount of extinction in the bulge (tracing
the column density of the cold interstellar medium) is also 
strongly linked to black hole growth and the age of the bulge stars.

\item
At fixed central stellar velocity dispersion,  we find that the radial
distribution of the stellar mass in the 
host galaxy shows only small ($\sim$ 10\%)  variations
as a function of black hole growth rate and 
the colour of the disk.

\end {itemize}

The suggested scenario is one in which the source of gas that builds the
bulge and black hole is a relatively low mass reservoir of cold gas in the disk. The
presence of the gas is necessary, but not sufficient for bulge and black hole growth.
Some mechanism must transport this gas inward in a time variable way. 
The duty cycle for such inflow periods (once cold gas in the outer disk is present)
must be relatively large, because a significant fraction  of the
galaxies in our sample have both blue outer disks and star formation in the bulge.  
After the gas has been transported into the bulge, the relative timescales
for bulge and black hole  growth must be also similar, because there
is a reasonably tight  correlation
between D4000 and L[OIII]/$M_{BH}$.

The disk gas itself is likely to be the result of the accretion of gas
from an external source. It is possible that these disks represent the
repository of baryonic material that has managed to condense out
of cooling flows in the centres of dark matter halos. 
As this gas is converted into stars in the bulge and disk, galaxies will turn
red in NUV-$r$. Further inflow can bring them back into the blue NUV-$r$
sequence. 
One question that arises is
whether the evolution of the disk  occurs through secular processes, i.e. 
the required rearrangement of angular momentum and mass  is achieved through 
collective phenomena such as bars, oval disks, spiral structure
and triaxial dark matter halos (see Kormendy \& Kennicutt 2004
for a recent review) , or whether the evolution is dominated
by more violent and rapid processes, such as mergers and strong
interactions. 
The similarity of the average stellar mass profiles of the quiescent
and active galaxies implies that the processes that lead to episodic growth
of the bulge and black hole do not usually involve major changes in the
galaxy structure.
We will address these issues in future work.

In summary, the  GALEX data presented in this paper have provided us 
with a new view of the bulge-dominated  galaxy population in which star 
formation has largely shut down by the present day.
We have learned that the transition from actively star-forming
galaxies at low stellar  surface mass densities to ``passive''
galaxies at high stellar surface densities does not imply that
disk formation has ceased altogether, but that disks have become
subdominant. In galaxies with large                
central velocity dispersions, the disks are inconspicuous in
the optical. However, star-forming outer disks 
still dominate the energy output of the
galaxy at ultraviolet wavelengths and this leads to a very
different accounting of which galaxies in the local Universe are
truly ``red and dead''. We  have also seen that these UV disks contain
the fuel for continued growth of  central supermassive black holes.
Understanding the processes by which UV-bright disks form
and why their formation is increasingly inhibited as the central velocity
dispersion of the bulge increases will be an important next step
in piecing together the puzzle of how massive galaxies came to be.

 
\acknowledgments
 
GALEX (Galaxy Evolution Explorer) is a NASA Small Explorer, launched in April 2003.
We gratefully acknowledge NASA's support for construction, operation, and science
analysis for the GALEX mission, developed in cooperation with the Centre
National d'Etudes Spatiales (CNES) of France and the Korean Ministry of Science
and Technology.

Funding for the creation and distribution of the SDSS Archive has been
provided by the Alfred P. Sloan Foundation, the Participating Institutions,
the National Aeronautics and Space Administration,
the National Science Foundation, the U.S. Department of Energy,
the Japanese Monbukagakusho, and the Max Planck Society.
The SDSS Web site is http://www.sdss.org/.
The SDSS is managed by the Astrophysical Research Consortium (ARC)
for the Participating Institutions. The Participating Institutions
are The University of Chicago, Fermilab, the Institute for Advanced Study,
the Japan Participation Group, The Johns Hopkins University,
the Korean Scientist Group, Los Alamos National Laboratory,
the Max-Planck-Institute for Astronomy (MPIA),
the Max-Planck-Institute for Astrophysics (MPA),
New Mexico State University, University of Pittsburgh,
University of Portsmouth, Princeton University,
the United States Naval Observatory, and the University of Washington.

\clearpage 
\begin{figure}
\epsscale{.80}
\plotone{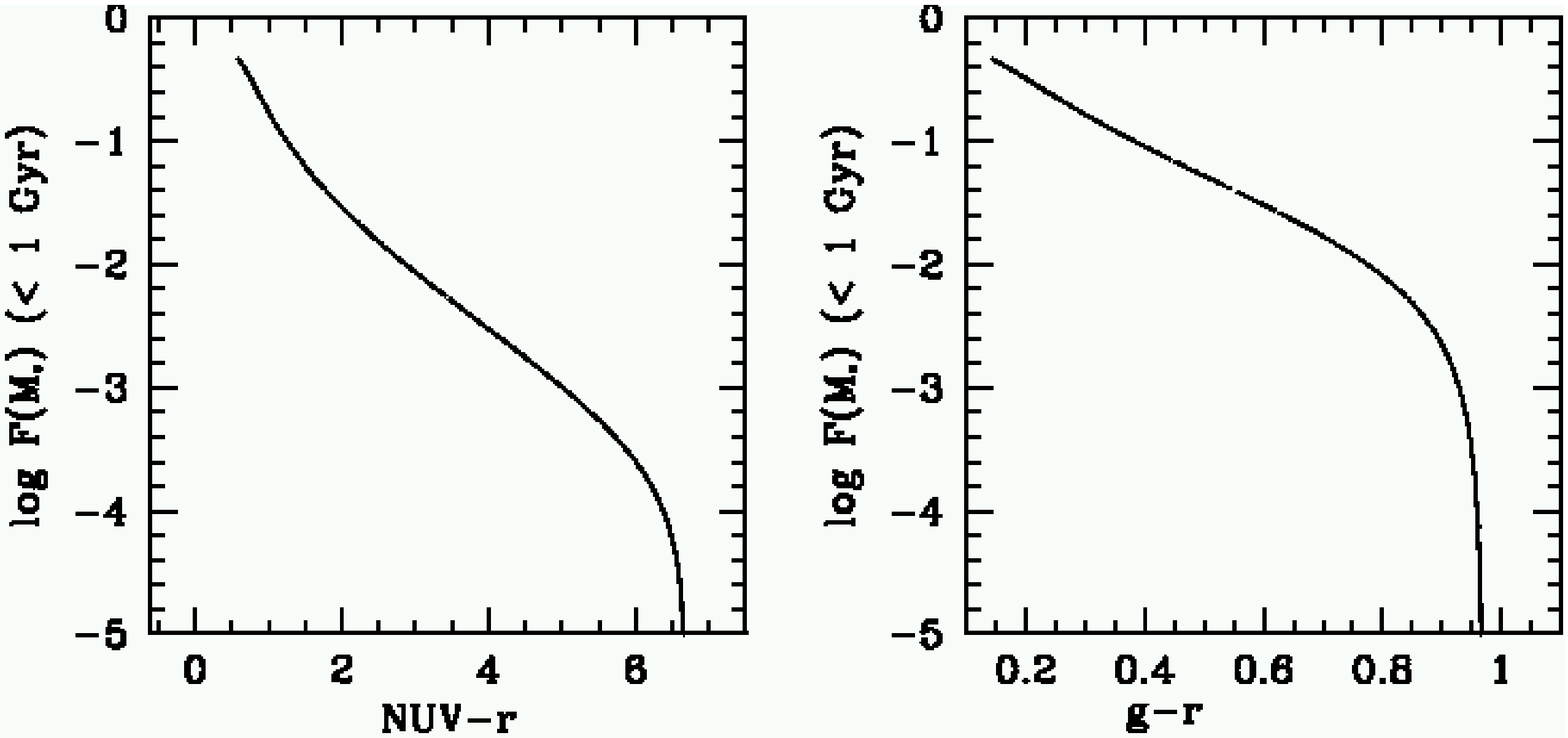}
\caption{The fraction of mass in stars in the galaxy formed in the last Gigayear is
plotted as a function of NUV-$r$ (left) and $g-r$ colour (right) for model
galaxies that gave smooth, exponentially declining star formation histories.}
\end{figure}

\begin{figure}
\epsscale{.80}
\plotone{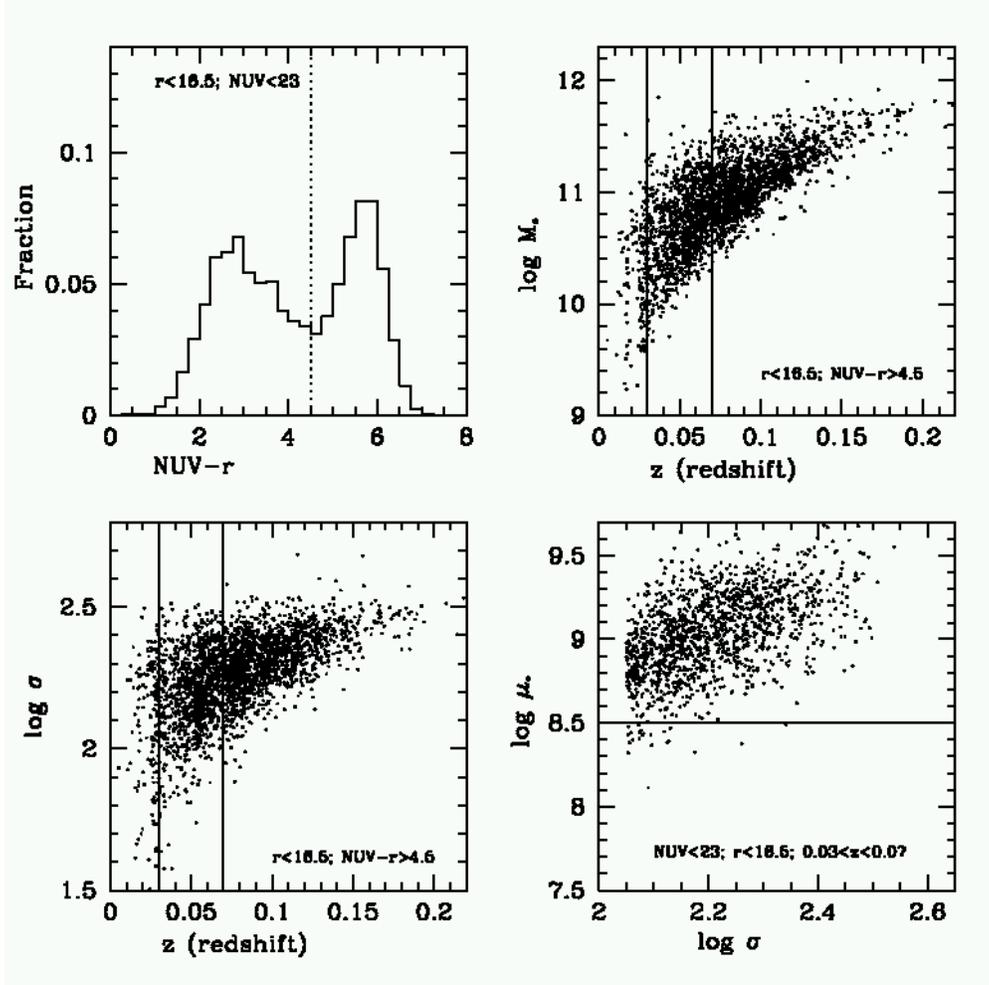}
\caption {
{\em Top left:} The NUV-$r$ colour distribution of galaxies with             
with $r< 16.5$ and NUV$< 23$ in our sample.
{\em Top right, bottom left:} Stellar mass and stellar velocity dispersion
is plotted as a function of redshift for galaxies with
NUV-$r > 4.5$, $r < 16.5$ and NUV $< 23$. {\em Bottom right:}  
Stellar surface mass density is plotted as a function of stellar velocity
dispersion for galaxies with  $\log M_* > 10.4$, $\log \sigma > 2.05$,  $r< 16.5$,
NUV$<23$ and $0.03 < z < 0.07$. The horizontal line indicates the ``transition''
value of $\mu_*$ between galaxies with ongoing star formation and
galaxies where star formation has largely shut down. } 
\end{figure}

\begin{figure}
\epsscale{.80}
\plotone{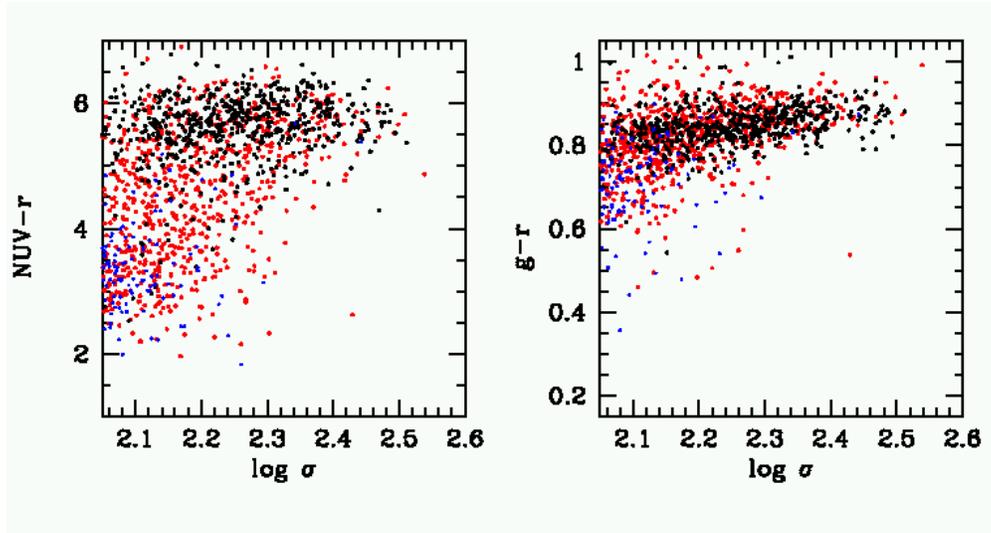}
\caption {
The relations between NUV-$r$ and $g-r$ colours and central stellar
velocity dispersion for the galaxies in our sample.
Black points indicate galaxies with emission lines that are too weak
to classify; red are AGN; blue are star-forming galaxies.}
\end{figure}

\begin{figure}
\epsscale{1.10}
\plotone{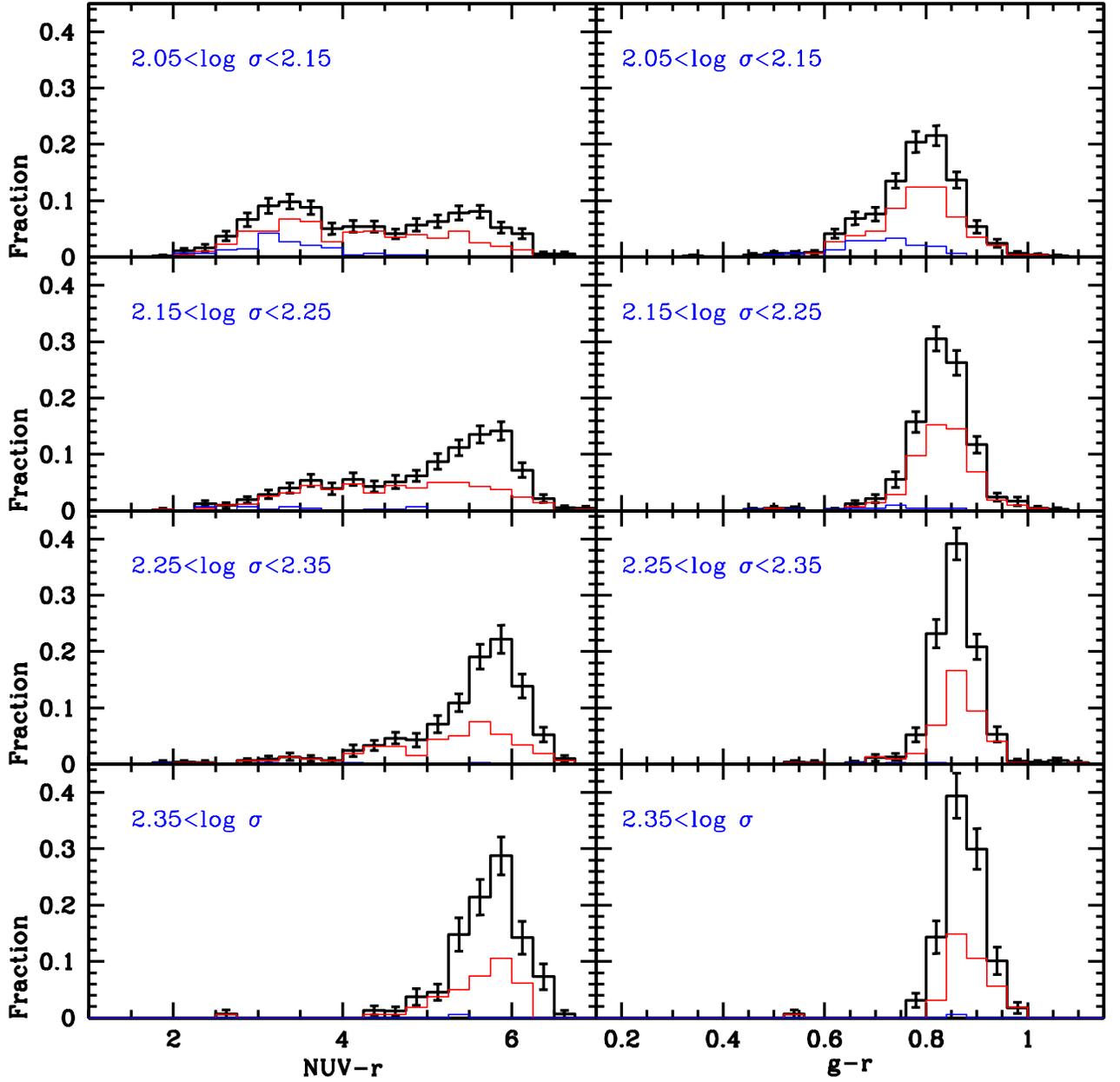}
\caption {
The distribution of NUV-$r$ and $g-r$ colours in different ranges of central
velocity dispersion.
Black histograms  are for the whole sample;                                
red shows the contribution from  AGN and  blue from  star-forming galaxies.}
\end{figure}

\begin{figure}
\epsscale{.80}
\plotone{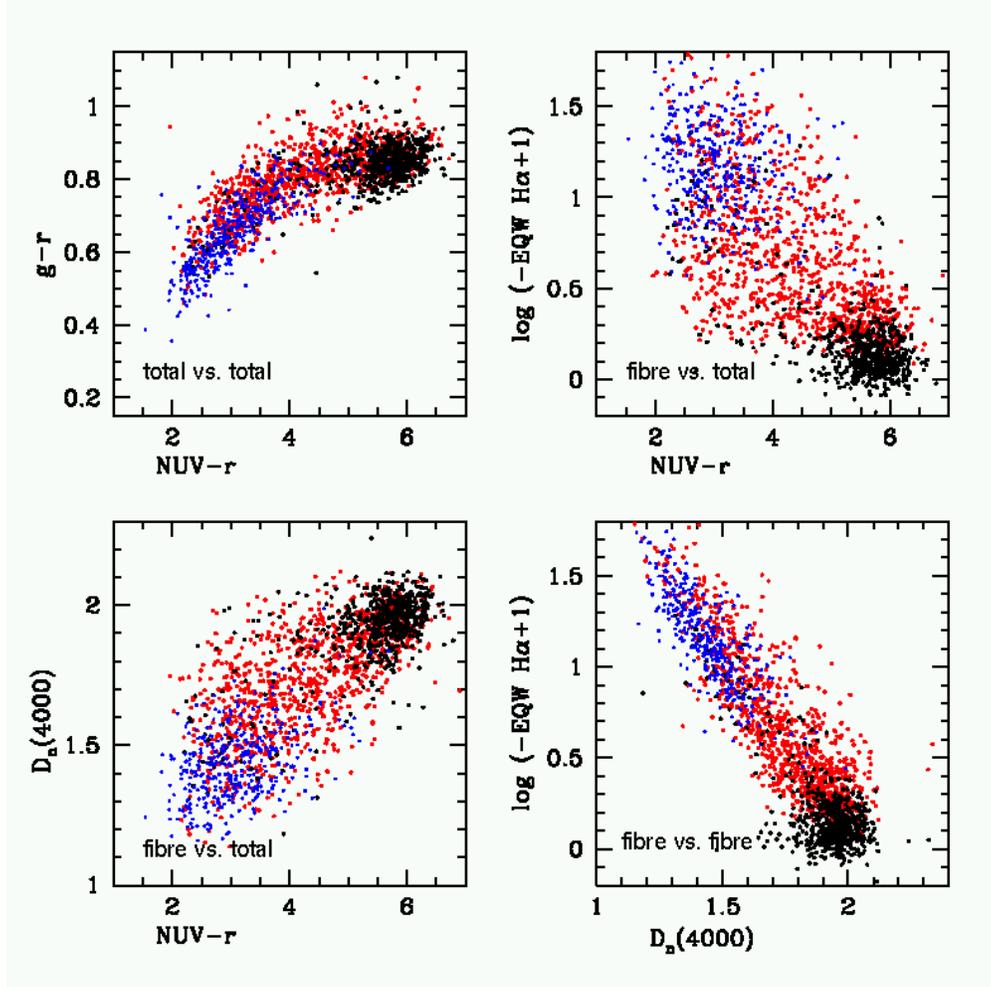}
\caption {
The relations between four different stellar population
indicators: the NUV-$r$ colour, the $g-r$ color,
the 4000 \AA\ break index D$_n$(4000),  and the equivalent
width of the H$\alpha$ emission line. Galaxies with emission
lines that are too weak to classify are shown as black points;
AGN are in red; star-forming galaxies are plotted in blue.}
\end{figure}

\begin{figure}
\epsscale{.83}
\plotone{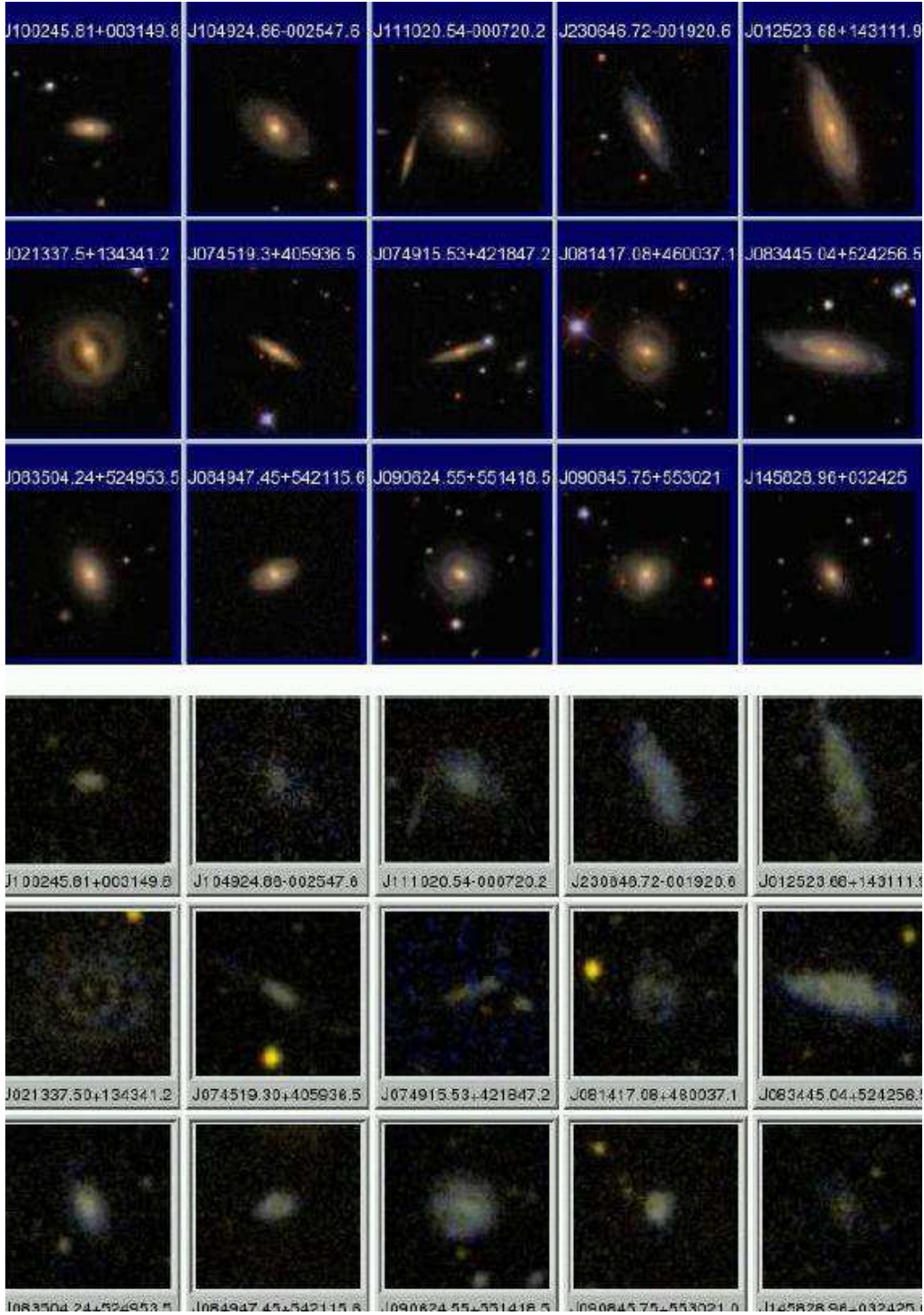}
\caption {
SDSS (top)  and GALEX (bottom) images  
 of UV-bright galaxies with low central H$\alpha$
equivalent widths. Each postage stamp image is 100 arcseconds on
a side, which correpsonds to a physical scale of
$\sim$ 100 kpc at these redshifts ($z=0.04-0.06$). 
Note that the UV disks are sometimes nearly as large as the
entire field! The SDSS fibre diameter is 3 arcseconds.}  
\end{figure}

\begin{figure}
\epsscale{.90}
\plotone{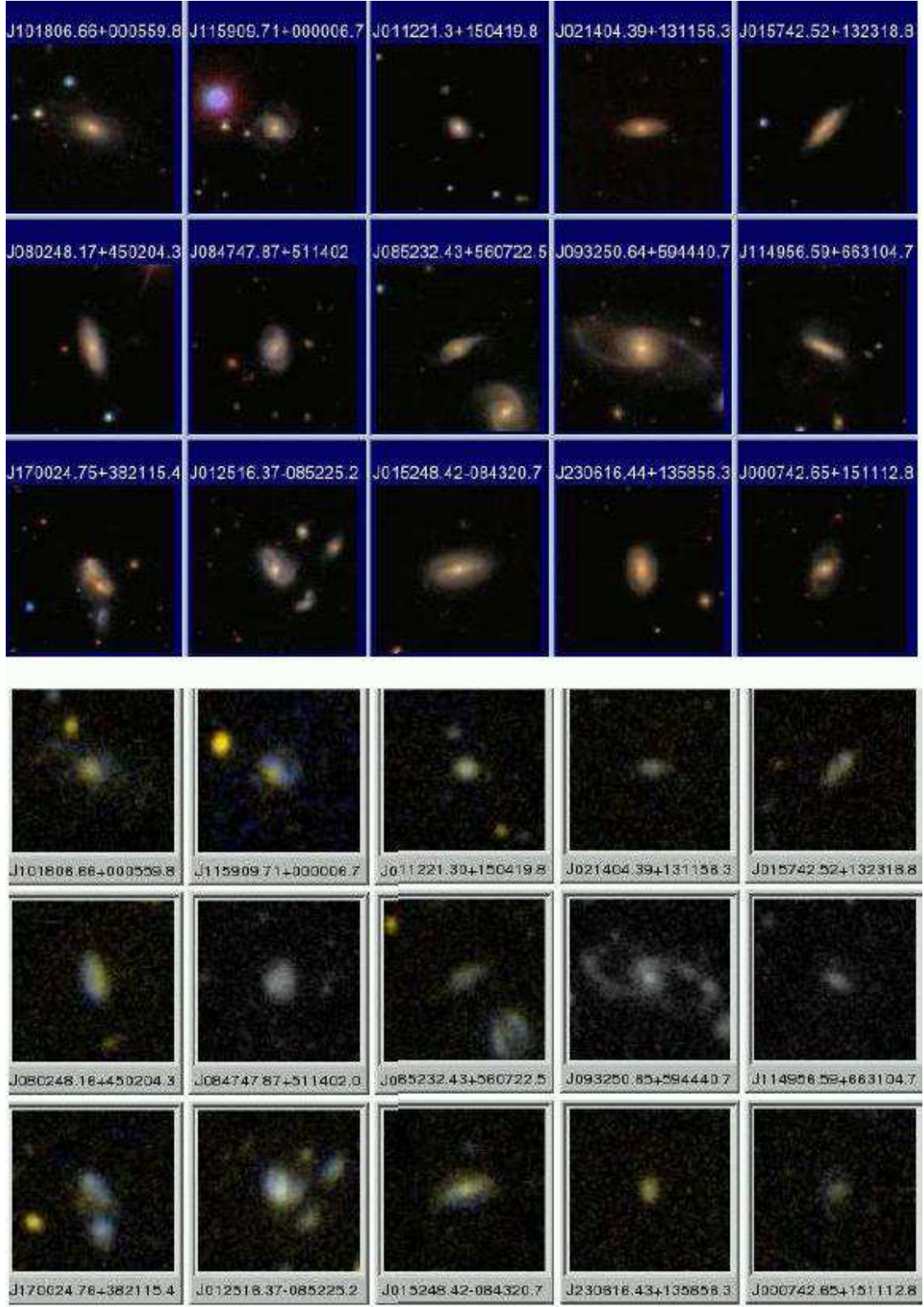}
\caption {
SDSS (top) and GALEX (bottom) images  
of UV-bright galaxies with high  central H$\alpha$
equivalent widths.}
\end{figure}

\begin{figure}
\epsscale{.90}
\plotone{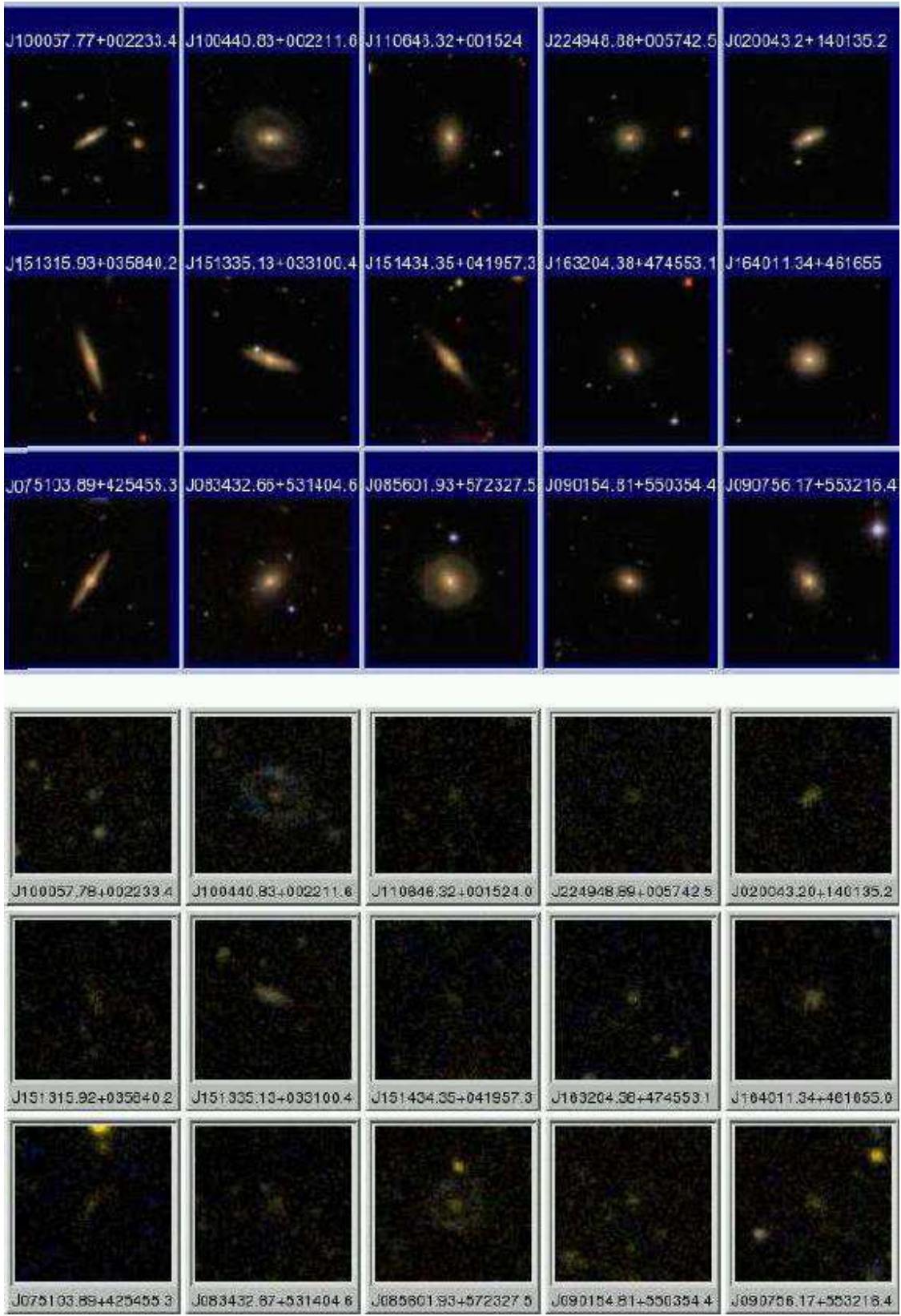}
\caption {
SDSS (top)  and GALEX (bottom) images  
of UV-faint galaxies.} 
\end{figure}

\begin{figure}
\epsscale{.80}
\plotone{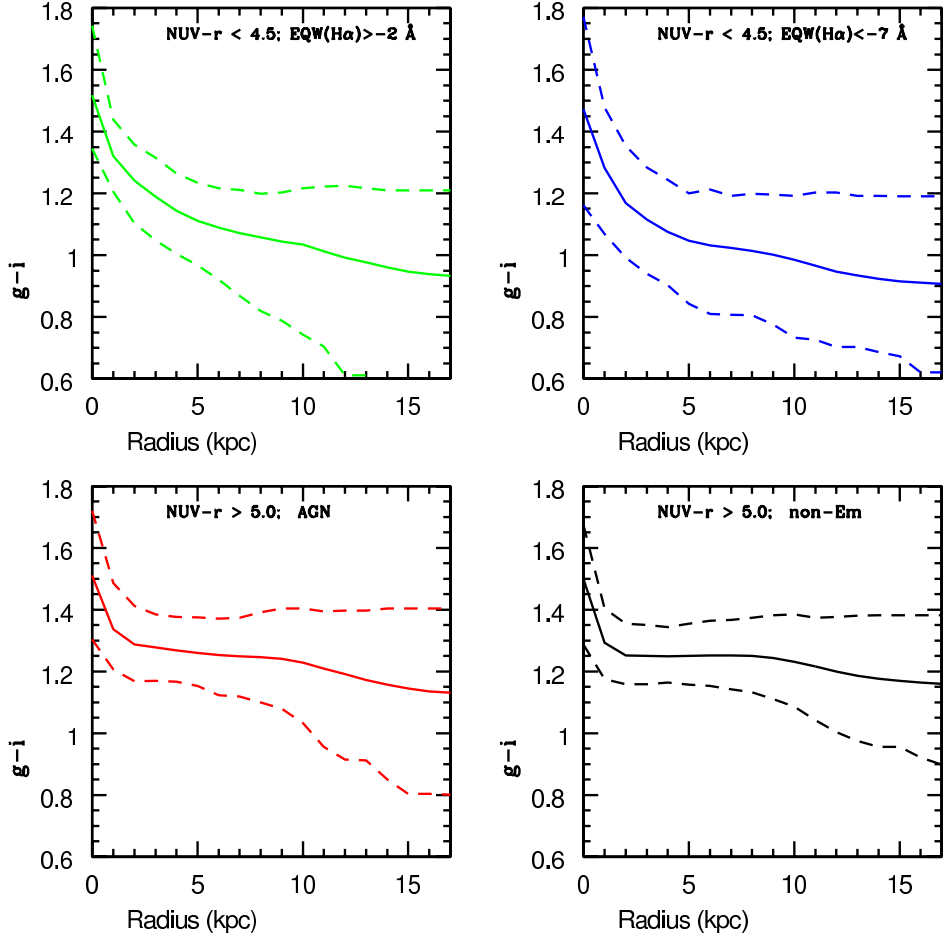}
\caption {
$g-i$ colour is plotted  as a function
of physical radius for UV-bright (upper panels) and UV-faint
(lower panels) galaxies with stellar velocity dispersions in
the range $2.05 < \log \sigma < 2.25$. The average profile
is shown as a solid line; the dashed lines indicate the 10-90
percentiles in the range of colour spanned by
the galaxies at a given radius.}
\end{figure}

\begin{figure}
\epsscale{.80}
\plotone{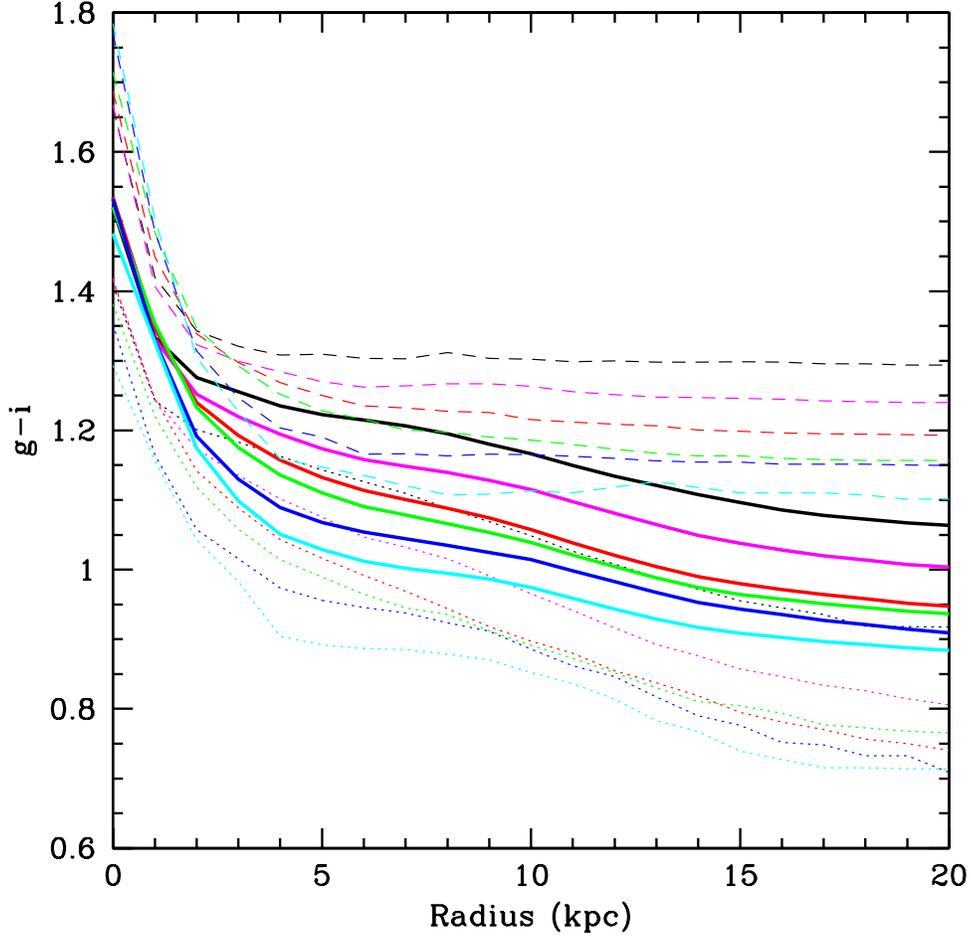}
\caption {
$g-i$ colour is plotted  as a function
of physical radius for galaxies in 6 different ranges of black hole
accretion rate, which is parametrized by
the quantity  $\log$ L[OIII]/M$_{BH}$. The black, magenta,red,green blue and
cyan lines are for AGN with  $\log$ L[OIII]/$M_{BH}$ in the range
[$< -2.0$];[ -2, -1.3]; [-1.3,- 0.6]; 
[-0.6, 0.1] ; [0.1, 0.7]; [$> 0.7$].  The solid lines show the average
profile, while the dashed and dotted lines indicate the 90th and 10th percentiles
of the colour distribution at a given radius.}
\end{figure}

\begin{figure}
\epsscale{.80}
\plotone{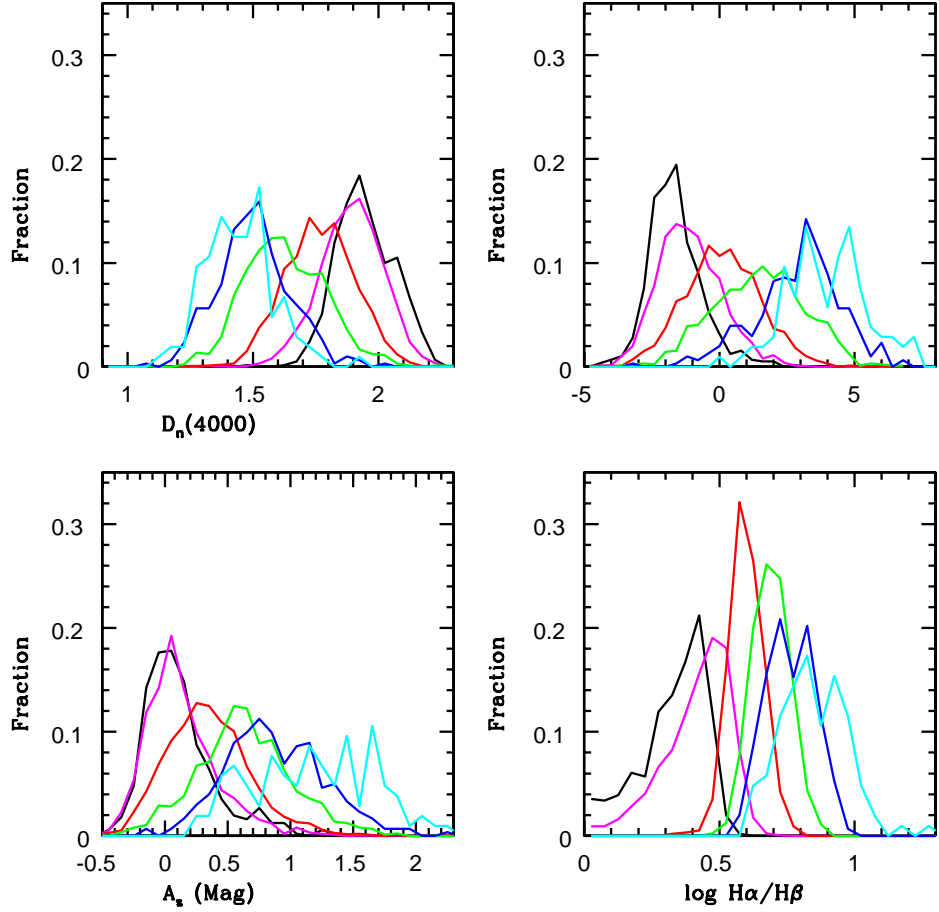}
\caption {
The distributions of  the age-sensitive spectral indices D$_n$(4000)
and H$\delta_A$, as well as two dust-sensitive measures, in bins
of black hole accretion rate parametrized by 
the quantity  $\log$ L[OIII]/M$_{BH}$. The colour-coding on the curves
is the same as in the previous figure.}
\end{figure}

\begin{figure}
\epsscale{.80}
\plotone{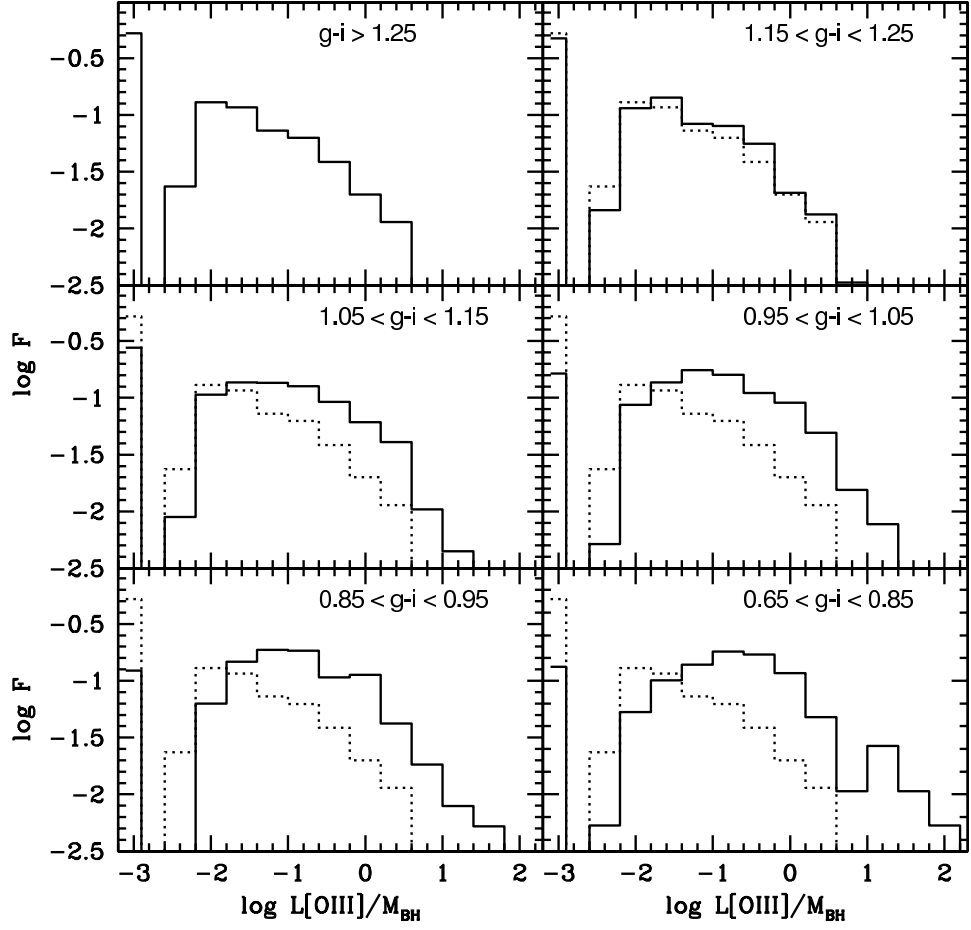}
\caption {
The distribution of $\log$ L[OIII]/M$_{BH}$ in bins of outer $g-i$ colour.
To guide the eye, the distribution shown in the first panel is repeated as a dotted line
in each subsequent panel.}
\end{figure}

\begin{figure}
\epsscale{.80}
\plotone{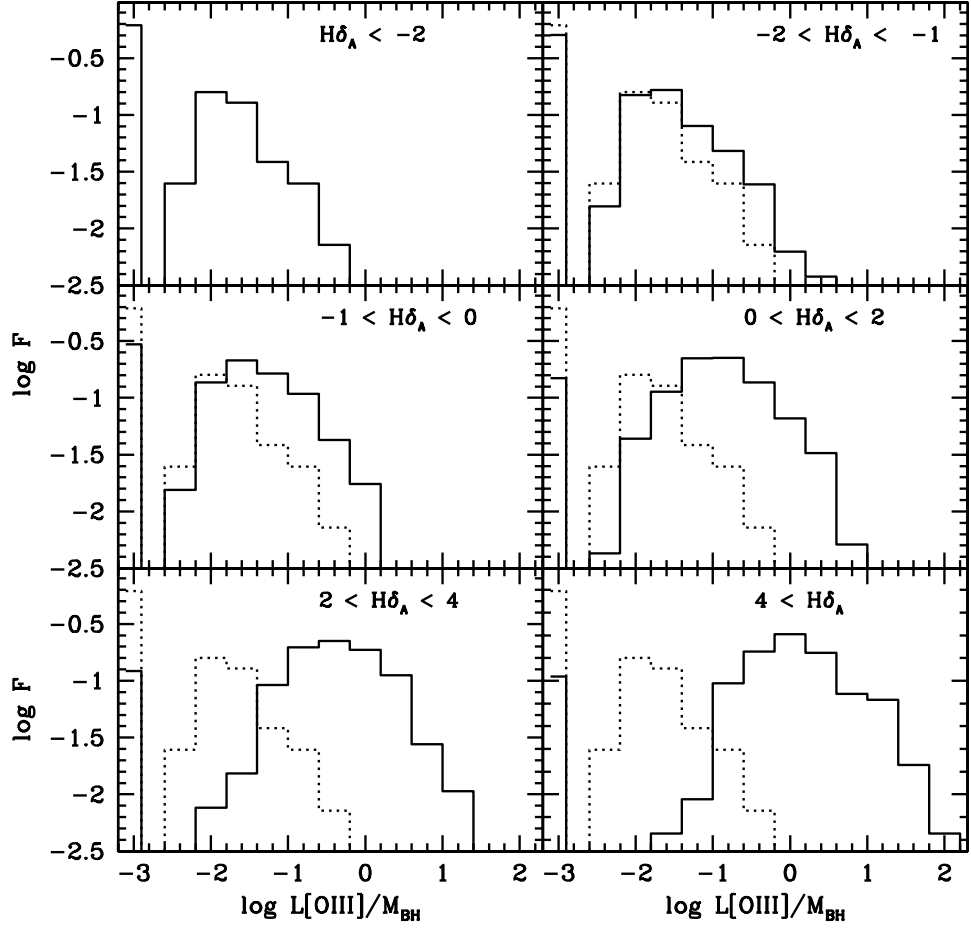}
\caption {
As in the previous figure,
except that the distribution of $\log$ L[OIII]/M$_{BH}$ 
is plotted in bins of H$\delta_A$  measured within
the fiber.}
\end{figure}

\begin{figure}
\epsscale{1.00}
\plotone{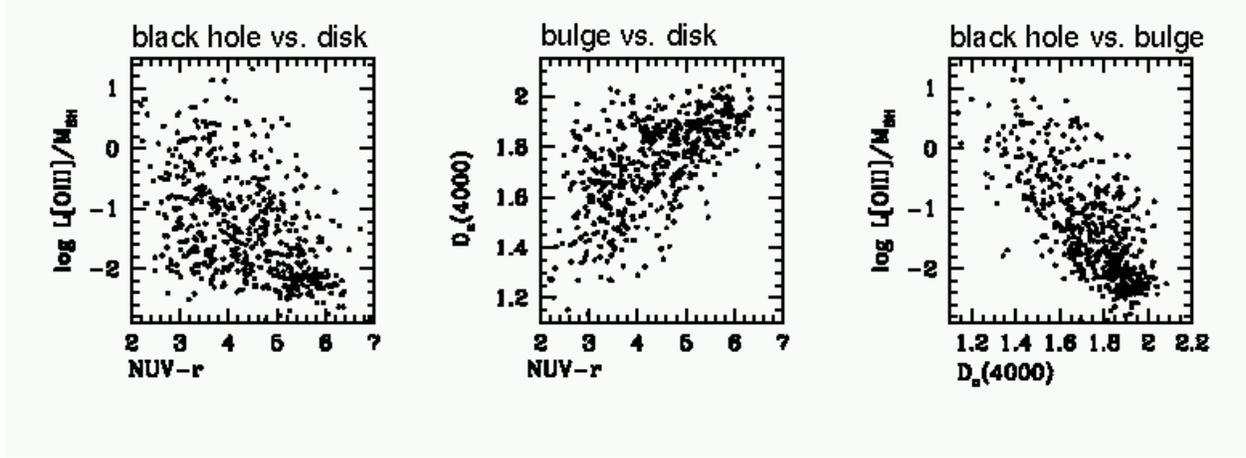}
\caption {
Correlations between  NUV-r (an indicator of the age of stars in the outer           
galaxy), D$_n$(4000) (an indicator of stellar age in the inner galaxy), and 
L[OIII]/$M_{BH}$ (an indicator of accretion rate onto the black hole).}
\end{figure}

\begin{figure}
\epsscale{.50}
\plotone{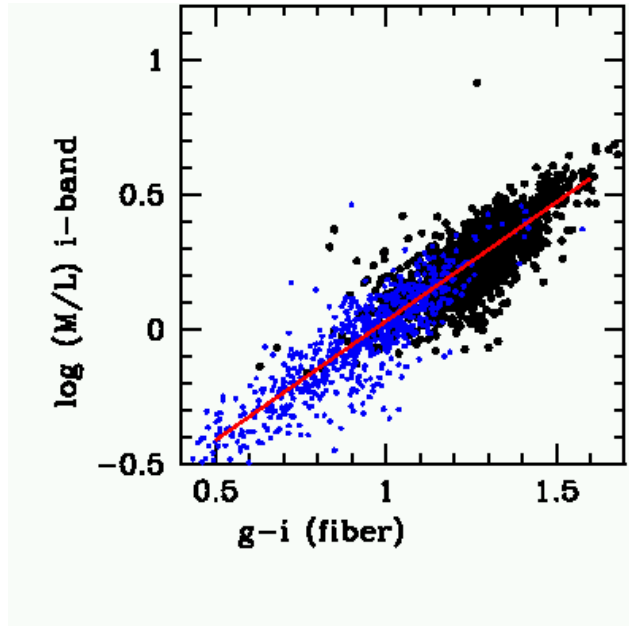}
\caption {
The $i$-band mass-to-light ratio estimated using stellar
absorption line indices is plotted as a function of $g-i$ fiber colour.} 
\end{figure}

\begin{figure}
\epsscale{1.00}
\plotone{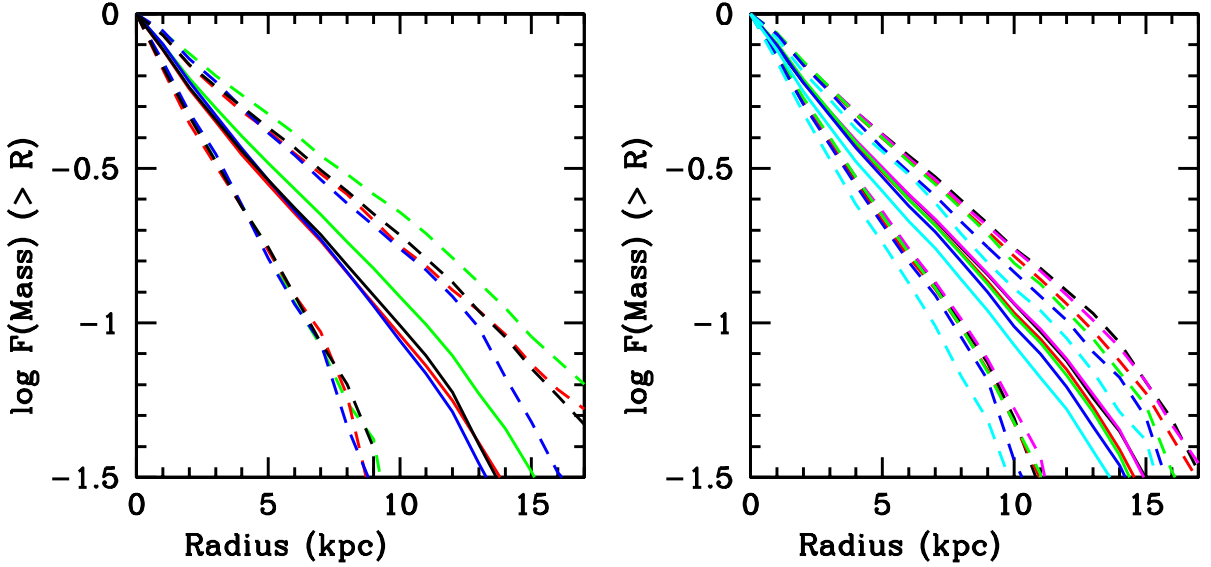}
\caption {
{\em Left:} The fraction of stellar mass exterior to  a given radius
is plotted as a function of radius.  The definition of the                    
coloured lines is the same as in Figure 9. The solid lines show 
the average stellar mass profiles. The dashed lines indicate the
lower 10th and upper 90th percentiles of the external mass at
a given value of R. UV-bright galaxies with little star
formation in their bulges have the most extended stellar
mass profiles. {\em Right:} Same as the left panel, except
that the definition of the coloured lines is the same as
in Figure 10. Galaxies with the highest values of
L[OIII]/M$_{BH}$ have the most concentrated stellar mass profiles. }   
\end{figure}

 
 
 





\end{document}